\documentclass[prl,twocolumn,showpacs,superscriptaddress]{revtex4}
\usepackage{amssymb}
\usepackage{graphicx}
\usepackage{amsmath}
\usepackage{mathrsfs}
\usepackage{epsfig}

\newcommand{\beq}{\begin{equation}}
\newcommand{\eeq}{\end{equation}}

\newcommand{\bea}{\begin{eqnarray}}
\newcommand{\eea}{\end{eqnarray}}

\begin{document}

\title{Looking into DNA breathing dynamics via quantum physics}
\author{Lian-Ao Wu}
\affiliation{Department of Theoretical Physics and History of
Science, The Basque Country University (EHU/UPV), PO Box 644, 48080
Bilbao, Spain}
\author{Stephen S. Wu}
\affiliation{Department of Life Science, Queen's University, Kingston, Ontario K7L 3N6, Canada}
\author{Dvira Segal}
\affiliation{Department of Chemistry and Center for Quantum Information and Quantum
Control, University of Toronto, 80 St. George Street, Toronto, Ontario M5S
3H6, Canada}
\date{\today}

\begin{abstract}
We study generic aspects of bubble dynamics in DNA under time dependent
perturbations, for example temperature change,
by mapping the associated Fokker-Planck equation to a quantum {\it
time-dependent} Schr\"{o}dinger equation with imaginary time. In the
static case we show that the eigenequation is exactly the same as
that of the $\beta$-deformed nuclear liquid drop model, without the
issue of non-integer angular momentum. A universal breathing dynamics is demonstrated
by using an approximate method in quantum mechanics. The calculated bubble
autocorrelation function qualitatively agrees with experimental
data. Under time dependent modulations, utilizing the adiabatic
approximation, bubble properties  reveal memory effects.
\end{abstract}

\pacs{05.40.-a,02.50.-r, 87.14.gk, 87.10.Mn}

\maketitle


\emph{Introduction.} The stability of the double helix structure of
DNA can be attributed to the phosphodiester bonds in the single
stranded sugar backbone and hydrogen bonds between complementary
base pairs of opposite strands. However, the hydrogen bonds between
parallel strands can be locally broken under physiological conditions preceding
events such as DNA replication, transcription, denaturation and protein
binding \cite{Bio}. A change in environmental conditions such as \textit{p}H or
temperature may provide the energy required to {\it progressively} open the
hydrogen bonds, producing domains of single-stranded DNA (bubbles).
Eventually, e.g. upon heating, denaturation occurs and the two strands separate altogether.
Understanding the underlying mechanisms behind breathing fluctuations \cite{breathing} and
force-assisted denaturation \cite{forced},
may provide further insights onto DNA structure and function. 

Breathing dynamics was recently detected through fluorescence fluctuations in a tagged double stranded DNA
\cite{Oleg}.
Various treatments were employed for simulating this effect:
the master-equation approach \cite{Metzler-master},
stochastic dynamic simulations of the Peyrard-Bishop-Dauxois model
\cite{Bishop}, and by 
adopting the Poland-Scheraga free energy function \cite{Poland66}, solving
the associated Fokker-Planck equation \cite{Hanke03}. Specifically,
it has been suggested that thermally induced breathing processes could
be mapped into the quantum Coulomb problem, with non-integer orbital
angular momentum \cite{Metzler-col}. Here the temperature, a
parameter in the free energy, plays a role in distinguishing
repulsive from attractive Coulombic potentials.

In this letter we are concerned with DNA bubble dynamics when
temperature, or other control parameter, varies in time. Adopting a
generic unpairing energy function, we study the bubble survival
behavior based on the mapping of the Fokker-Planck equation with
{\it time-dependent parameters}, to the quantum
\emph{time-dependent} Schr\"{o}dinger equation with imaginary time.
By employing approximate quantum mechanics methods,
a universal breathing dynamics is demonstrated, insensitive to the details
of the free energy function.
Moreover, we exemplify memory effects 
when external parameters (e.g. temperature or {\it p}H) are
slowly varied.

\emph{Model.}
The Poland-Scheraga free energy for a single bubble
can be written as \cite{Poland66,Hanke03}
\begin{equation}
F(x)=\Gamma (x)+ck_B T\ln (x+1)+\gamma_0, \label{eq:Free}
\end{equation}
with $x\geq 0$ as the bubble size in units of base pairs.
$\gamma_{0}$,  the free energy barrier to form the initial bubble, is next
omitted as it only introduces a constant shift in energy.
The entropy loss associated with the formation of a
closed polymer ring is incorporated by the factor $ck_BT\ln (x+1)$,
whereas $\Gamma(x)=2k_BT\int^{x}\varepsilon (y)dy$
represents the free energy for the dissociation
of $x$ base pairs \cite{Gevorkian08,commentT}, $k_B$ is the Boltzmann constant and $T$ is the temperature.
 The function $\Gamma (x)$ or $\varepsilon (y)$ may be modeled based on
experimental data. A simple model \cite{Hanke03} assumes that
$\Gamma (x)=-\gamma_{1}\frac{\Delta T}{T_m}x$, where $\Delta
T=T-T_m$ with $T_m$ being the melting temperature and
$\gamma_{1}=4k_BT_{\gamma }$; $T_{\gamma}=37^{\circ}$C is the
reference temperature.
Since we are interested here in the time-evolution of the bubble
distribution due to a change in a parameter $\kappa$, e.g.
temperature or \textit{p}H, we write $\Gamma=\Gamma (\Delta \kappa
,x)$ where $\Delta \kappa =\kappa -\kappa _{c}$;  $\kappa_{c}$  the
critical value of $\kappa$. Note that $\Gamma$ should be an odd
function of $\Delta \kappa$.

At a finite temperature, the one dimensional bubble dynamics can be modeled using
the overdamped Langevin equation with a Gaussian white noise
\cite{Metzler-col}
\bea \dot x=-D\frac{\partial{ F}}{\partial x}  + \eta; \,\,\,
\langle \eta(t) \eta(\tau)\rangle =2 k_B T D \delta(t-\tau),
\label{eq:lang} \eea
where $D$ is a kinetic coefficient of units $(k_B T \times s)^{-1}$.
The corresponding probability density $P=P(x,t)$ satisfies
the Fokker-Planck equation \cite{Denisov08},
\begin{equation}
\frac{\partial P}{\partial t}=\frac{\partial }{\partial x} \left( f^{\prime} P \right)
+\frac{1}{2}\frac{\partial ^{2}P}{
\partial x^{2}},
\label{eq:Fokker}
\end{equation}
where $f^{\prime }=\partial f/\partial x$, and $f(x)=F(x)/2k_BT$. The time variable was redefined
 $2D k_B T t \rightarrow t$. Introducing a \emph{dressed}
transformation, $P=e^{- f(x)}\tilde{P}$= $e^{-\frac{\Gamma
(x)}{2k_BT}}(x+1)^{-\mu }\tilde{P}$;  $\mu=c/2$, leads to
\begin{equation}
-\frac{\partial \tilde{P}}{\partial t}= H\tilde{P}; \,\,\,\
H=
-\frac{1}{2}\frac{\partial
^{2}}{\partial x^{2}}+V(x,t),
\label{eq:schroedinger}
\end{equation}
with a time dependent potential energy
\begin{eqnarray}
V(x,t)=U(x)+\frac{\mu (\mu +1)}{2x^{2}}- \frac {\partial f}{\partial t},
\label{eq:potential0}
\end{eqnarray}
where we assumed that the time dependent parameters are $T$ and $\varepsilon$.
The potential $U(x)$ is  given by
\begin{equation}
U(x)=\frac{\varepsilon (x)^{2}}{2}+\frac{\mu \varepsilon (x)}{x}-\frac{
\varepsilon ^{\prime }(x)}{2},
\label{eq:potentialU}
\end{equation}
assuming that $x\gg 1$.  Eq. (\ref {eq:schroedinger}) resembles the
\emph{time-dependent} Schr\"{o}dinger equation with imaginary time
for a particle in a time-dependent potential. For the static case
the dynamics superficially resembles
 the radial equation of a particle in a central potential $U(x)$ with
centrifugal barrier $\mu (\mu +1)/2x^{2}$. However, in the
quantum-mechanical case the angular momentum $\mu$ must be an
integer. It is thus of fundamental interest to identify a quantum system which permits
real values for $\mu$.

\bigskip \emph{Nuclear Liquid Drop Model.}
The Bohr Hamiltonian \cite{Bohr75} in the nuclear liquid drop model
with a mass parameter $B_{2}=1$ is given by ($\hbar=1$)
\begin{equation}
H_{B}=-\frac{1}{2}\left[ \frac{1}{\beta ^{4}}\frac{\partial
}{\partial \beta }\beta ^{4}\frac{\partial }{\partial \beta
}-\frac{C_{5}(\gamma, \Omega )}{\beta ^{2}}\right] +V(\beta ,\gamma
).
\end{equation}
Here $\beta$ and $\gamma$ are the parameters corresponding to the
shape of a nucleus as an incompressible drop with quadrupole
deformation, $\Omega$ is the Euler angle onto the body-fixed axes,
and $C_{5}(\gamma, \Omega)$ is the Casimir operator of the SO(5)
group \cite{Wilets56}. For a family of potentials $V(\beta ,\gamma
)=U(\beta )+V(\gamma )/\beta^{2}$ \cite{Wu96}, the $\beta$ degree of
freedom can be separated,
\bea
&&\left[ -\frac{1}{2}\frac{\partial ^{2}}{\partial \beta ^{2}}+U(\beta )+\frac{
\mu (\mu +1)}{2\beta ^{2}}\right] u(\beta )=Eu(\beta ), \\
&&\lbrack C_{5}(\gamma, \Omega ) + 2 V(\gamma )]\varphi (\gamma,
\Omega )=[\mu (\mu +1)-2]\varphi (\gamma, \Omega ), \nonumber \eea
so as the Bohr Hamiltonian eigenstates are given by $u(\beta)\varphi
(\gamma, \Omega )/\beta ^{2}$. In a $\gamma$- unstable situation,
$V(\gamma )=0$, $\mu =1,2,...,$ are integers. However, in general
situations the effective Hamiltonian $H=-\frac{1}{2}\frac{\partial
^{2}}{\partial \beta ^{2}}+U(\beta )+\frac{\mu (\mu +1)}{2\beta
^{2}}$ has the exact same form as that of the breathing bubble
(\ref{eq:schroedinger}), with $\beta $ replaced by $x$, and $\mu$
any positive number. This suggests that
a nuclear liquid drop model, rather than a particle in a central potential \cite{Metzler-col},
better describes bubble dynamics in double-stranded polymers.

{\it Static limit.}  When all variables are time-independent
the probability density $\tilde{P}$ of (\ref{eq:schroedinger}) can be expanded
in the normalized eigenstates $\Psi _{n}$ solving
$H\Psi _{n}=E_{n}\Psi _{n}$,
\begin{equation}
P(x,t)=e^{-f(x)}\sum_{n}c_{n}e^{-E_{n}t}\Psi _{n}(x),
\label{eq:P(xt)}
\end{equation}
where the coefficients $c_{n}$ are determined by the initial
condition and the completeness of $\Psi_{n}$. We specify next the
boundary conditions and distinguish between scattering potentials
and binding potentials. To account for
bubble closure absorbing boundary conditions are taken for vanishing
bubble size, $\Psi _{n}(0)=0$. Likewise, for considering a complete
denaturation of a long strand with a maximum bubble size $L$,
the absorbing condition $\Psi_{n}(L\rightarrow \infty)=0$ is implied.
In order to satisfy both conditions, the family of
functions $\varepsilon (x)$ should be monotonic for large
$x$ values so that $V(x)$ is a binding potential. For instance, if $\varepsilon (x)$ is a
polynomial of degree $M>0$, the generated potential $V(x)$ [see Eqs.
(\ref{eq:potential0}) and (\ref{eq:potentialU})] is always a binding
potential with the asymptotic behavior $\frac{\mu (\mu +1)}{2x^{2}}
\stackrel {x \rightarrow 0}  {\longrightarrow} \infty $;
$\frac{\varepsilon (x)^{2}}{2} \stackrel {x \rightarrow \infty}
{\longrightarrow} \infty$. In contrast, if $M=0$, $\varepsilon (x)$
is a constant corresponding to the Coulomb's potential, and the
total potential is now a {\it scattering potential}, allowing the
function $\Psi _{n}(L\rightarrow \infty)$ to differ from zero.

\emph{WKB Analysis.}
When time approaches infinity the transition probability (\ref{eq:P(xt)}) reads
\begin{equation}
P(x,t)e^{E_{g}t}\approx c_{g}e^{-f(x)}\Psi _{g}(x),
\label{eq:Pg}
\end{equation}
where $\Psi_{g}(x)$ is the ground state of the given potential with
eigenenergy $E_g$. In the scattering case $\Psi_{g}(x)$ is an
oscillating function of $x$, while a bound ground state is usually
nodeless and localized at a certain region of $x$. What is the
effect of the factor $e^{-f(x)}$ on the dynamics? When
acting on the scattering ground state it affects the long time
behavior of the transition probability leading to closure or
denaturation of DNA bubbles  \cite{Metzler-col}. On the other hand,
a bound ground state  $\Psi_{g}(x)$ approaches zero when
$x\rightarrow \infty$, thus  the role of the
$e^{-f(x)}$ factor becomes influential. If the speed
of its divergence is slower than the convergence of $\Psi_{g}(x)$,
the bubble tends to close rather than to denaturate, and vice-versa.
Qualitative analysis can be made in terms of the traditional WKB approximation
\cite{WKB}. The exponential factor of the ground state is given by
$\Psi_{g}(x)\varpropto
 e^{ -\int^{x}dy\sqrt{2(V(y)-E_{g})} }$; $E_{g}<V(y)$.
In the asymptotic large $x$ limit,
the probability (\ref{eq:Pg}), omitting the time dependent part, reduces to
$P(x)\varpropto e^{ -\int^{x}dy\left[ \varepsilon
(y)+\sqrt{\varepsilon (y)^2-2E_{g}}\right]}.$
For the Coulomb potential, $\varepsilon(y)=\varepsilon_0$ is a constant,
therefore $P\varpropto  e^{ -\left(\varepsilon _{0}+\left|
\varepsilon _{0}\right| \frac{\mu }{\mu +1}\right)x }$ \cite{comment}.
More generally, for bound potentials
$V(x)\stackrel{x\rightarrow \infty}{\longrightarrow} \frac{\varepsilon(x)^2}{2}$, therefore
$\Psi_{g}(x\rightarrow \infty)\propto e^{-\int^{x}\left|
\varepsilon (y)\right| dy}$
yielding the probability distribution
\begin{equation}
P(x)\varpropto e^{-\int^{x}\left( \varepsilon (y)+\left| \varepsilon
(y)\right| \right) dy}. \label{eq:WKB}
\end{equation}
Since the integrand is
non-negative, $\int^{x}\left( \varepsilon (y)+\left| \varepsilon (y)\right|
\right)dy$ either increases for $\varepsilon(y)>0$, leading to bubble closure,
or does not change with $x$ for $\varepsilon(y)<0$,
so as the integrated probability linearly scales with size.
The WKB analysis thus provides a universal long time behavior,
insensitive to the details of the unpairing energy function.
However, the WKB method is usually not suitable for obtaining the exact functional behavior,
an example is provided below.

\emph{An exactly solvable example.}
The transition probability $P$
from an initial bubble of size $x_{0}$ to a bubble of final size $x$ at time
$t$ is given by (\ref{eq:P(xt)})
\bea P(x,x_{0},t)=
e^{-f(x)+f(x_0)}\sum_{n}e^{-E_{n}t}\Psi _{n}(x)\Psi _{n}(x_{0}),
\label{eq:dist} \eea
with the initial condition $P(x,x_{0},0)=\delta (x-x_{0})$.
At long times it is approximately given by
\bea P(x,x_{0},t)\stackrel {t \rightarrow \infty} {\longrightarrow}
e^{-f(x)+f(x_0)}
e^{-E_{g}t}\Psi _{g}(x)\Psi _{g}(x_{0}).
\label{eq:distg} \eea
In order to simplify our analysis,
we consider the following expansion for the unpairing function
$\varepsilon (x)=\varepsilon _{0}+2\varepsilon
_{1}x+O(x^{2})$. Truncating the series after
the linear term results in $\Gamma(x)=2k_BT(\varepsilon
_{0}x+\varepsilon _{1}x^{2})$, generating the potential
$U(x) =2\varepsilon _{1}^{2}x^{2}+2\varepsilon _{1}\varepsilon _{0}x+\frac{%
\mu \varepsilon _{0}}{x}
+\frac{\varepsilon _{0}^{2}+2\varepsilon _{1}(2\mu -1)}{2}$,
see Eq. (\ref{eq:potentialU}).
If $\varepsilon _{1}=0$, the potential reduces to the Coulomb potential as in \cite{Metzler-col}.
However, since the effect of $\varepsilon_1$ dominates at large distances,
one should consider its contribution,
for example, by using a perturbation series  \cite{Saad-GSHO}.
For simplicity we assume next that $\varepsilon_{0}=0$, resulting in
the spiked harmonic oscillator potential (\ref{eq:potential0})
\bea
V(x)=2\varepsilon _{1}^{2}x^{2}+\varepsilon _{1}(2\mu -1) + \frac{\mu (\mu
+1)}{2x^{2}},
\label{eq:model1}
\eea
with the exact ground state  \cite{Saad-SHO}
\bea
\Psi _{g}(x)=\left[ \frac{\sqrt{8\left| \varepsilon _{1}\right| }}{\tilde \Gamma
(\mu +\frac{3}{2})}\right] ^{\frac{1}{2}}(\sqrt{2\left| \varepsilon _{1}\right| }x)^{\mu
+1}e^{-\left| \varepsilon _{1}\right| x^{2}},
\label{eq:psiG}
\eea
where $E_g=(3+2\mu)|\varepsilon_1|+\varepsilon_1(2\mu-1)$ and $\tilde \Gamma(z)$
is the Gamma function.
We substitute Eq.
(\ref{eq:psiG}) into (\ref{eq:distg}) and obtain
\bea P(x,x_{0},t\rightarrow \infty)& \approx & \frac{2(2\left|
\varepsilon _{1}\right| )^{\mu +\frac{3}{2}}\left( x_{0}\right) ^{2\mu
+1}}{\tilde\Gamma (\mu +\frac{3}{2})}
\nonumber\\
&\times &x\frac{e^{-(\varepsilon _{1}+\left| \varepsilon _{1}\right| )x^{2}}}{
e^{-(\varepsilon _{1}-\left| \varepsilon _{1}\right| )x_{0}^{2}}}e^{-E_{g}t}.
\label{eq:SHO}
\eea
When $\varepsilon _{1}>0$, the distribution is localized near $x=0$,
implying bubble closure. In contrast,  for $\varepsilon_{1}<0$ the
distribution leans towards larger $x$ values, $P\propto x$. The WKB
approximation (\ref{eq:WKB}) thus produced the correct exponential
factor, 
but it could not provide the factor $x$.
The correlation function $C(t)$, proportional to the integrated
survival probability $C(t)\propto \int_{0}^{L}P(x,x_0,t)dx$; $L$ is
the length of the DNA chain, can be recorded
experimentally \cite{Oleg}. We explore next this quantity as well as  the
first passage time distribution $W(t)=-dC(t)/dt$.

\begin{figure}
{\hbox{\epsfxsize=85mm \epsffile{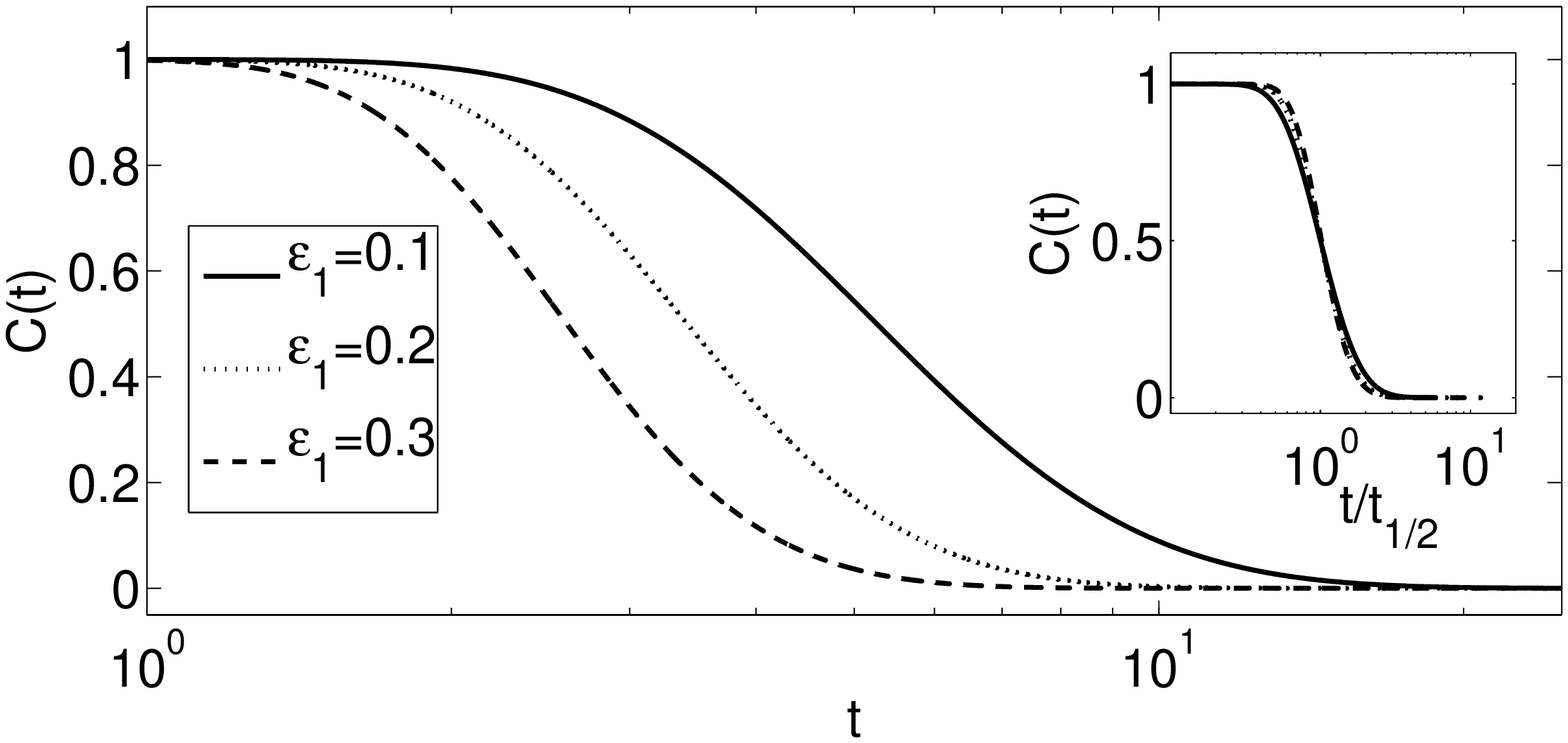}}}
\caption{Autocorrelation function at $\mu$=1/2
and $x_0$=5  [Eq. (\ref{eq:closeC})]
for $\varepsilon_1$=0.1 (full); $\varepsilon_1=0.2$ (dotted) and $\varepsilon_1=0.3$ (dashed).
(inset) The curves with rescaled times $t\rightarrow t/t_{1/2}$.
}
\label{Fig1}
\end{figure}

{\it Results for $\varepsilon_{1}>0$.}
Using the sum (\ref{eq:dist}) we obtain a superposition of exponentially decaying functions,
corresponding to various relaxation modes \cite{Oleg},
\bea C(t) =\frac{(2 \varepsilon _{1} )^{\mu +\frac{1}{2}}\ x_{0} ^{2\mu
+1}}{\tilde \Gamma (\mu +\frac{3}{2})(\mu +\frac{1}{2})^{-1}}
\label{correlation}
\sum_{n=0}^{\infty}\frac{\xi ^{n+\mu +\frac{1}{2}}L_{n}^{\mu +\frac{1}{2}}(2
\varepsilon _{1} x_{0}^{2})}{n+\mu +\frac{1}{2}}. \nonumber\\ \label{eq:Cg}
\eea
Here $\xi \equiv e^{-4 \varepsilon_{1}t}$, and
$L_{n}^{\mu+\frac{1}{2}}$ is the associated Laguerre Polynomial.
%
The first passage time distribution could be exactly calculated,
taking the time derivative of this expression. At $\mu =1/2$ it has
the following form
\begin{equation}
W_{\mu=\frac{1}{2}}(t)=8(\varepsilon_{1}x_{0})^{2}\frac{e^{-4
\varepsilon_{1} t}\exp \left[\frac{2\varepsilon_{1}
x_{0}^{2}}{1-\exp (4 \varepsilon_{1} t)}\right]}{\left[1-\exp (-4
\varepsilon_{1} t)\right]^{2}},
\label{eq:closeW}
\end{equation}
with $W(0)=W(\infty)=0$, and a maximum in between,
resulting in a profile similar to that obtained in
\cite{Metzler-col}. The correlation function at $\mu=1/2$ is given
by
\begin{equation}
C_{\mu=\frac{1}{2}}(t)\propto 1-\exp \left[\frac{2\varepsilon_{1}
x_{0}^{2}}{1-\exp (4\varepsilon_{1} t)}\right],
 \label{eq:closeC}
\end{equation}
with the long time limit $C_{\mu=\frac{1}{2}}(t)\propto e^{-4
\varepsilon_{1}t}$. The bubble lifetime is therefore given by
$\tau_c=$ $1/4\varepsilon_1$, or
$\tau_c=[4\varepsilon_1(\mu+\frac{1}{2})]^{-1}$ in general cases,
see (\ref{eq:Cg}). On the other hand, at short times
$C_{\mu=\frac{1}{2}}(t)\propto 1-\exp(x_0^2/2t)$.
%
Fig. \ref{Fig1} presents the correlation function using the
analytical form $\varepsilon(x)=2\varepsilon_1 x$ for the unpairing
free energy, and $\mu=1/2$, see (\ref{eq:Free}). Notice that the
curves at different $\varepsilon_1$, corresponding e.g. to different
temperatures or DNA structures, follow the same universal temporal
behavior. When presented as a function of a rescaled time
[$t\rightarrow t/t_{\frac{1}{2}}$, where
$C(t_{\frac{1}{2}})=\frac{1}{2}$], the plots collapse into a single
curve, in a good agreement with experiments \cite{Oleg} and
other theoretical treatments \cite{Katz,Metzler-master}.
Incorporating $\varepsilon_0$ should result in a similar behavior.

{\it Results at $\varepsilon_{1}<0$.} In this case the DNA
fully denatures at long times, and correlations diverge.
At $\mu=1/2$ we can exactly obtain the first
passage time distribution
\begin{equation}
W_{\mu=\frac{1}{2}}(t)=\frac{8(\varepsilon_{1}x_{0})^{2}}{e^{4\left|
\varepsilon_{1}\right| t}-1} \left[ 1- e^{ -\frac{2\left|
\varepsilon_{1}\right| L^{2}}{\exp (4\left| \varepsilon_{1}\right|
t)-1} }\right],
\end{equation}
and the corresponding correlation function $C(t)$.
%
%
At long times both scale as $L^{2}$.

\emph{Time dependent effects.}
%
The adiabatic approximation is standardly applied to describe the
dynamic of systems under slowly varying time dependent Hamiltonians \cite{WKB}.
Since the relaxation time of the bubble, order of $\mu s$ \cite{Oleg}, is typically shorter
than the modulation time of a parameter $\kappa$, e.g.
the temperature, the quantum adiabatic approximation may be applied
to describe the dynamics in the imaginary-time Schr\"{o}dinger
equation (\ref{eq:schroedinger}).
Defining an instantaneous basis of eigenenergies $H(t)\left|
n(t)\right\rangle =E_{n}(t)\left| n(t)\right\rangle$, we obtain
 $\langle n|\dot k\rangle = \frac{\left\langle n\right| \overset{\cdot }{H}\left| k\right\rangle }{ \omega _{kn}}$,
where $\omega _{kn}(t)=E_{k}(t)-E_{n}(t)$. In the axial
representation the wave function is written as $\left| \Psi
(t)\right\rangle =\sum_{n}a_{n}(t)e^{-\int_{0}^{t}d\tau E_{n}(\tau
)}\left| n(t)\right\rangle$. Substituting this into
the imaginary-time Schr\"{o}dinger equation we get
\bea
\dot{a}_n=-a_n\langle n|\dot{n} \rangle
-\sum_{k\neq n}a_k(t)\frac{\left\langle n\right|
\dot {H}\left| k\right\rangle }{\omega _{kn}}e^{-\int_{0}^{t}d\tau \omega
_{kn}(\tau )}.
\eea
Under the adiabatic approximation the coefficients $a_{n}(t)$ evolve
independently from each other  since couplings between
states are negligible \cite{WKB}. In the present case we require that $\left|
\frac{\left\langle n\right| \dot H\left| k\right\rangle }{\omega
_{kn}}\right| e^{-\int_{0}^{t}d\tau \omega _{kn}(\tau )}\ll 1$. If
$\omega_{kn}(t)>0$, the exponential factor 
is always less than 1, while for $\omega_{kn}<0$ it may diverge at
long times. Therefore, the applicability of the adiabatic
approximation may be questionable for general instantaneous states \cite{Sarandy05},
yet for the ground state it is valid as long as the standard
adiabatic condition $\left| \frac{\left\langle n\right|
\overset{\cdot }{H}\left| k\right\rangle }{\omega_{kn}}\right| \ll
1$ holds. Under the adiabatic approximation the ground state
amplitude evolves according to $\dot a_{g}\approx -a_g\left\langle
g| \dot g\right\rangle$. However, since $\langle n|\overset{\cdot
}{n}\rangle $ is zero for any one-dimensional real wave function,
the overall function propagates as $\Psi(x,t)\sim \Psi_g(x,t)
e^{-\int_0^t d\tau E_g(\tau)}$ with $\Psi_g(x,t)$ as the instantaneous
solution (\ref{eq:psiG}).
Consider for example the potential $V(x,t)=(2\varepsilon _{1}^{2}-
\dot \varepsilon_{1})x^{2}+\varepsilon _{1}(2\mu -1)+\frac{\mu (\mu
+1)}{2x^{2}}$ [see Eqs.
(\ref{eq:schroedinger})-(\ref{eq:potentialU}) and
(\ref{eq:model1})], which has analytical instantaneous eigenstates.
To simplify, we further assume that the system initially occupies
the ground state of the potential $V(x,t=0)$. Under the adiabatic
approximation
\begin{equation}
P(x,t)\approx \frac{ \left( 2\sigma -2\varepsilon _{1}\right)
^{\frac{\mu}{2} +\frac{3}{4}}}{\sqrt{\tilde \Gamma \left(\mu
+\frac{3}{2}\right)/2}}xe^{-\sigma x^{2}}e^{-\int_{0}^{t}d\tau
E_{g}(\tau )},
 \label{eq:time}
\end{equation}
with the width parameter $\sigma
=\varepsilon_1+(\varepsilon_1^{2}-\frac{\dot
\varepsilon_1}{2})^{1/2}$, and $E_{g}(\tau)=\sigma(\tau)(2\mu
+3)-4\varepsilon_1(\tau)$. Rich information can  be obtained due
to the time-dependent evolution of $\varepsilon_1$. First, both the
width of the distribution and the peak position depend on
$\dot\varepsilon_1$, the rate at which the external parameters (e.g.
temperature) is changed.
Secondly, the processes of increasing and decreasing the control
parameter may reach the same value $\varepsilon_{1}$, yet they may
result in different shapes of the bubble distribution. Specifically,
the correlation function $C(t)\propto \left( \sigma -\varepsilon
_1\right)^{\frac{\mu} {2}+\frac{3}{4}} (1-e^{-\sigma
L^2})e^{-\int_{0}^{t}d\tau E_{g}(\tau)}$ includes the decay factor
$e^{-\int_{0}^{t}d\tau E_{g}(\tau)}$ which memorizes the different
pathways that $\varepsilon _{1}(t)$ undergoes. For example, the two
paths $\varepsilon _{1}(t)=1+t/100$ and
$\varepsilon_{1}(t)=1+t^{2}/100$ attain the same value at $t=1$, yet
the values of $\int_{0}^{t}d\tau E_{g}(\tau )$ are obviously
different, yielding distinct characteristic decay times. We expect
that this theoretical result could be observed experimentally.

\emph{Summary.} The dynamics of a single DNA bubble under time
dependent perturbations was studied by mapping the associated
Fokker-Planck equation to a quantum \emph{time-dependent}
Schr\"{o}dinger equation with imaginary time.
%
For a generic unbinding free energy function we analyzed bubble
breathing by using the WKB approximation, observing a universal
behavior. Specifically, a spiked harmonic oscillator potential
yielded results in qualitative agreement with experimental data.
Under slow time dependent modulations of e.g., the temperature or
{\it p}H, bubble dynamics reflects memory effects.

L. A. Wu has been supported by the Ikerbasque foundation. D. Segal acknowledges
the University of Toronto Start-up grant.



\begin{thebibliography}{99}
\bibitem{commentT}
For convenience,  temperature appears as a prefactor in  $\Gamma$.
In comparison to \cite{Gevorkian08},
our $\varepsilon(x)$   compensates this factor.
\bibitem{comment}
Our sign notation with respect to attractive/repulsive potentials is opposite to
\cite{Metzler-col}.
\end{thebibliography}
\end{document}